\documentstyle[aps,12pt]{revtex}

\begin{document}
\title{Theory of the solid-state physics on the turn: II. Importance of the
spin-orbit coupling for 3d-ion compounds: the case of NiO}
\author{R.J. Radwa\'{n}ski$^{a,b}$ and Z.~Ropka$^a$}
\address{$^a$Center for Solid State Physics, \'{s}w. Filip 5, 31-150 Krak\'{o}w\\
$^b$Inst. of Physics, Pedagogical University, 30-084 Krak\'{o}w.}
\maketitle

\begin{abstract}
The orbital and spin moment of the Ni$^{2+}$ ion in NiO has been calculated
at 0 K to be 0.54 $\mu _B$ and 1.99 $\mu _B$ respectively. Such large
orbital moment, more than 20 \% of the total moment of 2.53 $\mu _B$, proves
the need for the ''unquenching'' of the orbital moment in compounds
containing 3d ions. It turns out that the spin-orbit coupling is
indispensable for description of magnetic and electronic properties of
3d-ion compounds. These two findings, largely ignored at the nowadays
in-fashion solid-state thories, call for an advanced solid-state physics
theory.   
\end{abstract}


The discovery of the high-temperature superconductivity in the copper oxide
has revealed the enormous shortage of our general understanding of the
3d-ion compounds. Namely, the insulating state of La$_2$CuO$_4$ contradicts
the standard band-structure result that predicts it to be metal. This
dramatic breakdown of the ordinary band-electron theory has been known
already for years for 3d-ion monooxides [1-5]. Different reparations do not
lead to the consistent picture for 3d-ion systems known as Mott insulators.
At present after announcement of the essential importance of the single-ion
effects and many-electron discrete states in description of compounds
containing transition-metal atoms of the 4f [6] and 3d [7] groups the
fundamental controversy has become the scientific fact. In Ref.7 we have
argued that the spin-orbit coupling is essentially important (indispensable)
for the description of electronic and magnetic properties of compounds
containing 3d ions. The s-o coupling is largely ignored in the nowadays
in-fashion solid-state theories owing to the general conviction about the
quenching of the orbital moment and the weakness of the s-o coupling [1-5;
clearly admitted in Ref. 3 p. 7164]. Nowadays ionic systems like LaMnO$_3$,
LaCoO$_3$, MgV$_2$O$_4$ and NiO are in fashion. Here we will concentrate on
properties of NiO.

The aim of the present short paper is to report results of calculations of
the orbital and spin moment. In NiO the Ni ions are divalent. Their 6
electrons form the highly-correlated electron system with S=1 and L=3. The $%
^3$F term is 21-fold degenerated [7,8]. The Ni$^{2+}$ ion experiences the
crystal field, the spin-orbit coupling and the intersite spin-dependent
interactions. The latter term is approximated by the mean-field approach.
The fine electronic structure resulting from the cubic crystal-field
0interactions in the presence of the spin-orbit coupling, presented in Ref.
7, has the triply degenerated crystal-field ground state with the moments of
0 and $\pm $2.14 $\mu _B$.

We have calculated (the computer program is available on the written request
to the authors) the spin and orbital part of the magnetic moment as well as
their temperature dependence. The calculations have been performed by the
self-consistent way in order to get the magnetic state with Neel temperature
of 525 K. These calculations resemble much the calculations often perfomed
by us for rare-earth compounds [9]. The orbital and spin moment of the Ni$%
^{2+}$ ion in NiO has been calculated at 0 K to be 0.54 $\mu _B$ and 1.99 $%
\mu _B$ respectively. For calculations we have taken the spin-orbit coupling
constant $\lambda _{s-o}$=-41 meV and the octahedral crystal-field parameter
B$_4$=+2 meV (it yields the overall CEF splitting of order 2 eV). The
intersite spin-dependent interactions, expressed as the molecular field
acting on the Ni moment, amounts to 500 T.

The spin and orbital moments are parallel - the total moment amounts to 2.53 
$\mu _B$ at 0 K. With temperature all moments decrease vanishing at the
magnetic-ordering temperature. Such large orbital moment, more than 20 \% of
the total moment, proves that in compounds containing 3d ions the orbital
moment has to be ''unquenched''. We point out that the spin-orbit coupling
is fundamentally essential for the existence of this moment. Recent magnetic
x-ray experiments of Fernandez et al. has revealed the orbital moment of
0.34 $\mu _B$ at room temperature. We are are taking these values, 0.54 and
0.34 $\mu _B$, to be in nice agreement owing to the fact the orbital moment
becomes larger at low temperatures. In fact, we can say on basis of our
calculations that the spin-orbit coupling is indispensable for the
physically-adequate description of magnetic and electronic properties. These
two findings, the existence of the orbital moment and the importance of the
spin-orbit coupling, have been  largely ignored at the nowadays in-fashion
solid-state thories [1-5]. Obviously the present results call for more
advanced solid-state physics theory that will take these facts into account.
The need is unavoided as the orbital moment becomes revealed in modern
experiments [10].

In conclusion, the substantial orbital moment of 0.54 $\mu _B$ (at 0 K) in
NiO has been found in close agreement with the recent experimental finding.
Such large orbital moment, more than 20 \% of the total moment of 2.53 $\mu
_B$, proves that in compounds containing 3d ions the orbital moment has to
be ''unquenched''. It turns out that the spin-orbit coupling is
indispensable for description of magnetic and electronic properties of
3d-ion compounds.

** This work has been presented at the European Conf. Physics of Magnetism
99, held in Poznan (Poland, Scientific Chairman: R.Micnas and S.Krompiewski)
June 21-25, 1999 as the abstract P6.20. In the presentation it had two parts
with subtitles: I. Localized f electrons in heavy-fermion compounds: the
case of UPd2Al3 and II. Importance of the spin-orbit coupling for 3d-ion
compounds: the case of NiO. This paper has been submitted 15.05.1999 to the\
Org. Committee.

{\bf Figure Captions }

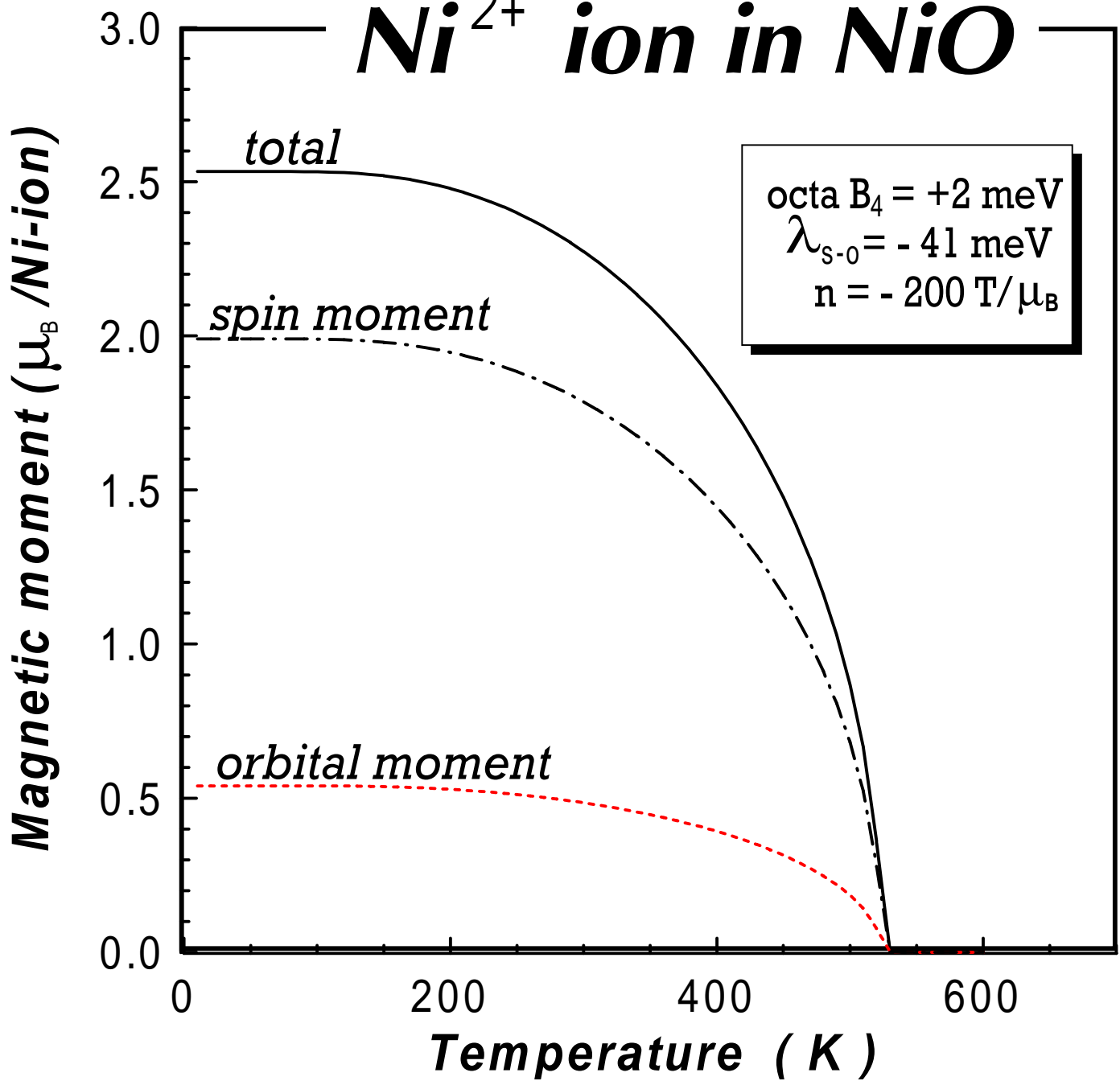
\end{document}